\documentclass[a4paper, 11pt]{article}
\usepackage[margin=1in]{geometry}

\usepackage[english]{babel}
\usepackage[utf8]{inputenc}
\usepackage{amsmath}
\usepackage{amssymb}
\usepackage{graphicx}
\usepackage{caption}
\usepackage{subcaption}
\usepackage{natbib}
\usepackage{xcolor}
\usepackage{float}

\title{Oracally Efficient Estimation of Functional-Coefficient Autoregressive Models}

\author{Qiwei Li\\University of California, Davis}

\date{November 25, 2014}

\begin{document}

\maketitle
\footnotetext{Research project for Spring and Summer 2014 as part of the Research Training Group. Supervised by Joshua Patrick.}

\section{Introduction}
\label{sec:intro}
Nonlinear autoregressive models is very useful for modeling many natural processes, however, the size of the class of these models is large.
Functional-coefficient autoregressive (FCAR) models are useful structures for reducing the size of the class of these models.
The FCAR model is defined as
\begin{equation}
X_{t}=\sum_{\alpha=1}^{p}m_{\alpha}\left(U_{t}\right)X_{t-\alpha}+\sigma\left(\textbf{X}_{t}\right)\varepsilon_{t},\qquad t=1,\ldots,n,
\label{eq:fcar}
\end{equation}
where $p$ and $d$ are positive integers, $m_{\alpha}\left(U_{t}\right)$
is a measurable function of the delay variable $U_t = X_{t-d}$,  for $\alpha=1,\ldots,p$,
$\sigma^{2}\left(\textbf{X}_{t}\right)$ is a variance function dependent on  $\textbf{X}_{t}=\left(X_{1},\ldots,X_{n}\right)^{\prime}$, and $\left\{ \varepsilon_{t}\right\} $ is a sequence of i.i.d.~random
variables with mean $0$ and variance $\sigma^{2}$. Although this structure reduces the class of nonlinear models, it is broad enough to include some common time series models as specific cases. Among these are the threshold autoregressive (TAR) model of \cite{tong1983}, the exponential autoregressive (EXPAR) model of \cite{haggan1981}, and the smooth transition autoregressive (STAR) model of  \cite{chan1986}.

\cite{chen1993} introduced the FCAR model and proposed a procedure for building the model based on arranged local regression which constructs estimators based on an iterative recursive formula that resembles local constant smoothing. \citet*{cai2000} used a local linear fitting method to estimate the coefficient functions. They used  the method on simulated data from an EXPAR model and assessed the fit by calculating the square root of  the average squared errors (RASE). In \cite{huang2004} a global smoothing procedure based on polynomial splines for estimating FCAR models is proposed. The authors  note that the spline method yields a fitted model with a parsimonious explicit expression which is an advantage over the local polynomial method. This feature allows one to produce multi-step ahead forecasts conveniently. Additionally, their spline method is less computationally intensive than the local polynomial method.

A recent development in estimating nonlinear time series data is the spline-backfitted kernel (SBK)  method of \cite{wang2007}. This method combines the computational speed of splines with the asymptotic properties of kernel smoothing. To estimate a component function in the model, all other component functions are ``pre-estimated" with splines and then the difference is taken of the observed time series and the pre-estimates. This difference is then used as pseudo-responses for which kernel smoothing is used to estimate the function of interest. By constructing the estimates in this way, the method does not suffer from the ``curse of dimensionality". 

In this paper, we adapt the SBK method to FCAR models. In section 2, the SBK methodology is discussed. In section 3, simulation results are used to show the oracle efficiencies of the method and we apply the method to real world data in section 4. We conclude with a discussion in section 5.

\section{Methodology}
\label{sec:method}
To motivate the estimation method for \eqref{eq:fcar}, we use the oracle smoothing
idea of \cite{linton1997} and \cite{wang2007}. Suppose we want to estimate $m_{\gamma}\left(U_{t}\right)$
in \eqref{eq:fcar}. If the coefficient functions $m_{\alpha}\left(U_{t}\right)$,
$\alpha=1,\ldots,p$, $\alpha\ne\gamma$, are known by ``oracle,''
then we can construct $\left\{ U_{t},X_{t-\gamma},Y_{\gamma,t}\right\} _{t=1}^{n}$,
where
\[
Y_{\gamma,t}=m_{\gamma}\left(U_{t}\right)X_{t-\gamma}+\sigma\left(\textbf{X}_{t}\right)\varepsilon_{t}=X_{t}-\sum_{\alpha=1,\alpha\ne\gamma}^{p}m_{\alpha}\left(U_{t}\right)X_{t-\alpha},
\]
from which we can estimate the only unknown function $m_{\gamma}\left(U_{t}\right)$.
This oracle smoother removes the ``curse of dimensionality'' since
there is only one unknown function to estimate. Clearly, the coefficent
functions, $m_{\alpha}\left(U_{t}\right)$, $\alpha=1,\ldots,p$,
$\alpha\ne\gamma$, are not known and must be estimated. For additive
models, \cite{linton1997} used marginal integration kernel estimates to estimate
the functions and \cite{wang2007} used an undersmoothed spline procedure. We
now adapt the procedure of \cite{wang2007} to estimate the FCAR model. 

We assume the delayed variable $U_{t}$ is distributed on the compact
interval $\left[a,b\right]$. Denote the knots as $a=\kappa_{0}<\kappa_{1}<\cdots<\kappa_{N}<\kappa_{N+1}=b$
where the number of interior knots are $N\sim n^{2/5}\ln n$. The
B spline basis functions are determined on the $N+1$ equally spaced
intervals with length $(b-a)\left(N+1\right)^{-1}$. The basis function
are defined as
\[
B_{J}\left(u\right)=\begin{cases}
1, & \qquad\kappa_{J}\le x<\kappa_{J+1},\\
0, & \qquad\text{otherwise,}
\end{cases}\qquad J=0,\ldots,N+1.
\]
The pre-estimates are defined as
\[
\hat{m}_{\alpha}\left(u\right)=\sum_{J=1}^{N+1}\hat{\lambda}_{\left(N+1\right)(\alpha-1)+J}B_{J}\left(u\right),\qquad\alpha=1,\ldots,p,
\]
where the coefficients $\left(\hat{\lambda}_{1},\ldots,\hat{\lambda}_{p\left(N+1\right)}\right)$
are solutions to the least squares problem
\begin{align}
\left\{ \hat{\lambda}_{1},\ldots, \right. & \left. \hat{\lambda}_{p\left(N+1\right)}\right\} = & \nonumber\\ &\underset{\mathbb{R}^{p\left(N+1\right)}}{\arg\min}\sum_{t=1}^{n}\left\{ X_{t}-\sum_{\alpha=1}^{p}\left(\sum_{J=1}^{N+1}\lambda_{\left(N+1\right)(\alpha-1)+J}B_{J}\left(U_{t}\right)\right)X_{t-\alpha}\right\} ^{2}.
\label{eq:leastsquare}
\end{align}
Define $\mathbf{X}_{t}=\left(X_{1},\ldots,X_{n}\right)^{\prime}$,
$\mathbf{X}_{\alpha}=\left(X_{1-\alpha},\ldots,X_{n-\alpha}\right)^{\prime},$
$\mathbf{U}_{t}=\left(U_{1},\ldots,U_{n}\right)^{\prime}$, $\hat{\boldsymbol{\lambda}}=\left(\hat{\lambda}_{1},\ldots,\hat{\lambda}_{p\left(N+1\right)}\right)^{\prime}$,
\[
\mathbf{B}=\left[\begin{array}{cccc}
B_{0}\left(U_{1}\right) & B_{1}\left(U_{1}\right) & \cdots & B_{N}\left(U_{1}\right)\\
B_{0}\left(U_{2}\right) & B_{1}\left(U_{2}\right) & \cdots & B_{N}\left(U_{2}\right)\\
\vdots & \vdots & \ddots & \vdots\\
B_{0}\left(U_{n}\right) & B_{1}\left(U_{n}\right) & \cdots & B_{N}\left(U_{n}\right)
\end{array}\right],
\]
and $\mathbf{Z}=\left(\mathbf{B}\circ\tilde{\mathbf{X}}_{1},\mathbf{B}\circ\tilde{\mathbf{X}}_{2},\cdots,\mathbf{B}\circ\tilde{\mathbf{X}}_{p}\right)$
where $\circ$ denotes the Hadamard product and $\tilde{\mathbf{X}}_{\alpha}$
is a $n\times\left(N+1\right)$ matrix with $\mathbf{X}_{\alpha}$
for each column. In matrix notation, the least squares estimates are
\[
\hat{\boldsymbol{\lambda}}=\left(\mathbf{Z}^{\prime}\mathbf{Z}\right)^{-1}\mathbf{Z}^{\prime}\mathbf{X}_{t},
\]
and the pre-estimates are 
\[
\hat{m}_{\alpha}\left(\mathbf{U}_{t}\right)=\mathbf{B}\left(\hat{\lambda}_{\left(N+1\right)\left(\alpha-1\right)},\ldots,\hat{\lambda}_{\alpha\left(N+1\right)-1}\right)^{\prime},\qquad\alpha=1,\ldots,p.
\]
We now define the ``pseudo-responses'' as
\[
\hat{Y}_{\gamma,t}=X_{t}-\sum_{\alpha=1,\alpha\ne\gamma}^{p}\hat{m}_{\alpha}\left(U_{t}\right)X_{t-\alpha},\qquad t=1,\ldots n.
\]
Define the vector of pseudo-responses as $\hat{\mathbf{Y}}_{\gamma}=\left(\hat{Y}_{\gamma,1},\ldots,\hat{Y}_{\gamma,n}\right)^{\prime}$.
The spline-backfitted kernel (SBK) estimate for the coefficient function
$m_{\gamma}\left(u\right)$ is 
\begin{equation}
\tilde{m}_{SBK,\gamma}\left(u\right)=\left(1,0\right)\left(\mathbf{V}^{\prime}\mathbf{W}\mathbf{V}\right)^{-1}\mathbf{V}^{\prime}\mathbf{W}\hat{\mathbf{Y}}_{\gamma},
\label{eq:sbll}
\end{equation}
where 
\[
\mathbf{V}=\left[\begin{array}{cc}
X_{p+1-\gamma} & X_{p+1-\gamma}\left(U_{p+1}-u\right)\\
\vdots & \vdots\\
X_{n-\gamma} & X_{n-\gamma}\left(U_{n}-u\right)
\end{array}\right],
\]
$\mathbf{W}=\text{diag}\left\{ K_{h}\left(U_{p+1}-u\right),\ldots,K_{h}\left(U_{n}-u\right)\right\} $,
\begin{equation*}
K_{h}\left(u\right)=h^{-1}\frac{15}{16}\left(1-\left(\frac{u}{h}\right)^{2}\right)^{^{2}}I_{\left\{ \left|u/h\right|\le1\right\} },
\label{eq:k}
\end{equation*}
$I_{\left\{ x\right\} }$ is an indicator variable equal to one if
$x$ and zero otherwise, and $h$ is a bandwidth selected by the rule of thumb criterion of \cite{fan1996}. 
Likewise, we define the oracle kernel smoother as
\begin{equation}
\tilde{m}_{O,\gamma}\left(u\right)=\left(\mathbf{V}^{\prime}\mathbf{W}\mathbf{V}\right)^{-1}\mathbf{V}^{\prime}\mathbf{W}\mathbf{Y}_{\gamma}.
\label{eq:o}
\end{equation}

\section{Simulation Results}
\label{sec:sims}
In this section, we present two simulation results on our finite-sample behavior of the SBK estimators of the functional-coefficient autoregressive model.
The dataset is generated from the FCAR model,
\[
X_{t}=\sum_{\alpha=1}^{p}m_{\alpha}\left(U_{t}\right)X_{t-\alpha}+\sigma\left(\textbf{X}_{t}\right)\varepsilon_{t},\qquad t=1,\ldots,n,
\]
The functional-coefficient term is set to be $m_{\alpha}\left(U_{t}\right)$ = $A_{\alpha} sin\left(\omega \pi U_{t}\right)$ for $\alpha = 1,...,p$, where $U_{t}=X_{t-d}$ is a delayed variable with $d = p+1$ and the predictor $X_{t}$ is generated from the distribution of $X_{t} \sim N(0,1)$.
We ran two sets of simulations, one with $p = 4$ and one with $p=10$. When $p=4$, we set $d = 5$, $A=\left(0.5,-0.5,0.5,-0.5\right)^{\prime}$, and $\omega = 4.5$. When $p = 10$, we set $d = 11$, $A=\left(0.5,-0.5,0.5,-0.5,0.5,-0.5,0.5,-0.5,0.5,-0.5\right)^{\prime}$, and $\omega = 1.5$. For all simulated models the error term is $\varepsilon_{t}\sim N\left(0,1\right)$ with
\[
\sigma\left(U_{t}\right) = 0.1 \left( \frac{\sqrt{p}}{2}\right) U_{t}
\frac{\left(5-e^{\sum_{i=1}^p{\frac{\left| X_{i} \right|}{p}}}\right)}{\left(5+e^{\sum_{i=1}^p{\frac{\left| X_{i} \right|}{p}}}\right)}
\]
which ensures heteroscedasticity with $\sigma \left(U_{t} \right)$ roughly proportional to dimension $p$.

To implement the SBK estimater, we choose spline degree to be 0 to ensure that we get  undersmooth pre-esitmates. The choice of number of knots $N_{n}$ for our spline estimator is,
\[
N_{n} = \min \left(\left \lfloor{ c_{1}n^{\frac{1}{4}}\log{n}}\right \rfloor + c_{2}, \left \lfloor{\frac{n}{2d}}\right \rfloor \right)
\]
where $c_{1}$ and $c_{2}$ are tuning constants. The choice of these constants $c_{1}$ and $c_{2}$ makes little difference for a large sample, thus we set $c_1=c_2=1$. In addition, we want $N_{n} \leq n/\left(2d\right)$. This ensures number of terms for the solution of the least squares problem \eqref{eq:leastsquare} is no greater than $n/2$, which is necessary when the sample size $n$ is moderate and dimension $p$ is high.  
Both SBK estimators $\tilde{m}_{SBK,\alpha}\left(u\right)$ and oracle smoother $\tilde{m}_{O,\alpha}\left(u\right)$ are obtained by local linear regression defined in \eqref{eq:sbll} and \eqref{eq:o} with the rule-of-thumb bandwidth.

We ran 500 replications for sample sizes $n = 100,500,1000,1500$. For both simulations, we chose to estimate the $\alpha_{1}$ and $\alpha_{4}$ components and compared the fit of our SBK estimators $\tilde{m}_{SBK,\alpha}\left(u\right)$ to the oracle smoothers $\tilde{m}_{O,\alpha}\left(u\right)$ by the relative efficiency, which is 
\begin{equation*}
\text{eff}_{\alpha} = \frac{\frac{1}{n}\sum ^{n}_{i=1}\left\{\tilde{m}_{SBK,\alpha}\left(u\right)-m_{\alpha}\left(u\right) \right\}^{2}}{\frac{1}{n}\sum ^{n}_{i=1}\left\{\tilde{m}_{O,\alpha}\left(u\right)-m_{\alpha}\left(u\right) \right\}^{2}}
\label{eq:eff}
\end{equation*}
Since for small sample sizes, the density of our relative efficiencies are skewed, we chose the mode, median, and variance to show the simulated results in Table \ref{table:efftable}.

\begin{table}[H]

\centering
\begin{tabular}{l l | ccc | ccc}
\hline
   &  & \multicolumn{3}{|c|}{eff$_{1}$} &  \multicolumn{3}{|c}{eff$_{4}$}     \\
$d$	&$n$	& mode      & median & variance & mode      & median & variance\\
\hline
4	& 100  & 0.274 & 0.510 & 1.009 & 0.210 & 0.530 & 1.429 \\
	& 500  & 0.722 & 0.776 & 0.634 & 0.524 & 0.730 & 0.905 \\
	& 1000 & 0.787 & 0.872 & 0.328 & 0.735 & 0.861 & 0.538 \\
	& 1500 & 0.857 & 0.890 & 0.365 & 0.710 & 0.831 & 1.172 \\
10	& 100  & 0.155 & 0.362 & 4.920 & 0.173 & 0.349 & 10.22 \\
	& 500  & 0.180 & 0.378 & 1.491 & 0.186 & 0.338 & 0.419 \\
	& 1000 & 0.328 & 0.500 & 1.555 & 0.236 & 0.408 & 1.150 \\
	& 1500 & 0.347 & 0.644 & 1.778 & 0.287 & 0.502 & 2.987 \\
\hline
\end{tabular}
\caption{Relative efficiency between $\tilde{m}_{SBK,\alpha}\left(u\right)$ and $\tilde{m}_{O,\alpha}\left(u\right)$.}
\label{table:efftable}
\end{table}

For both dimensions, the relative efficiencies are converging to 1 as the sample size increases. However, for high dimensions, the convergence is slower than low dimensions as expected. We also expect to see the variance of these relative efficiencies decrease when the sample size increases. It seems to be the case for low dimensions except the variance for $\alpha_{4}$ when $n=1500$ jumps back up by a large amount. The reason is one of the values is 18.11 which pulls up the variance. If this value is removed, the variance is then 0.589. Moreover, the efficiencies are not stable for high dimensions. The variance jumps from 10.22 to 0.419 when the sample increased from 100 to 500 and then goes back up to 2.987 when sample size is 1500. 

\begin{figure}[H]
\centering
\begin{subfigure}[b]{0.49\textwidth}
\centering
\includegraphics[width=\textwidth]{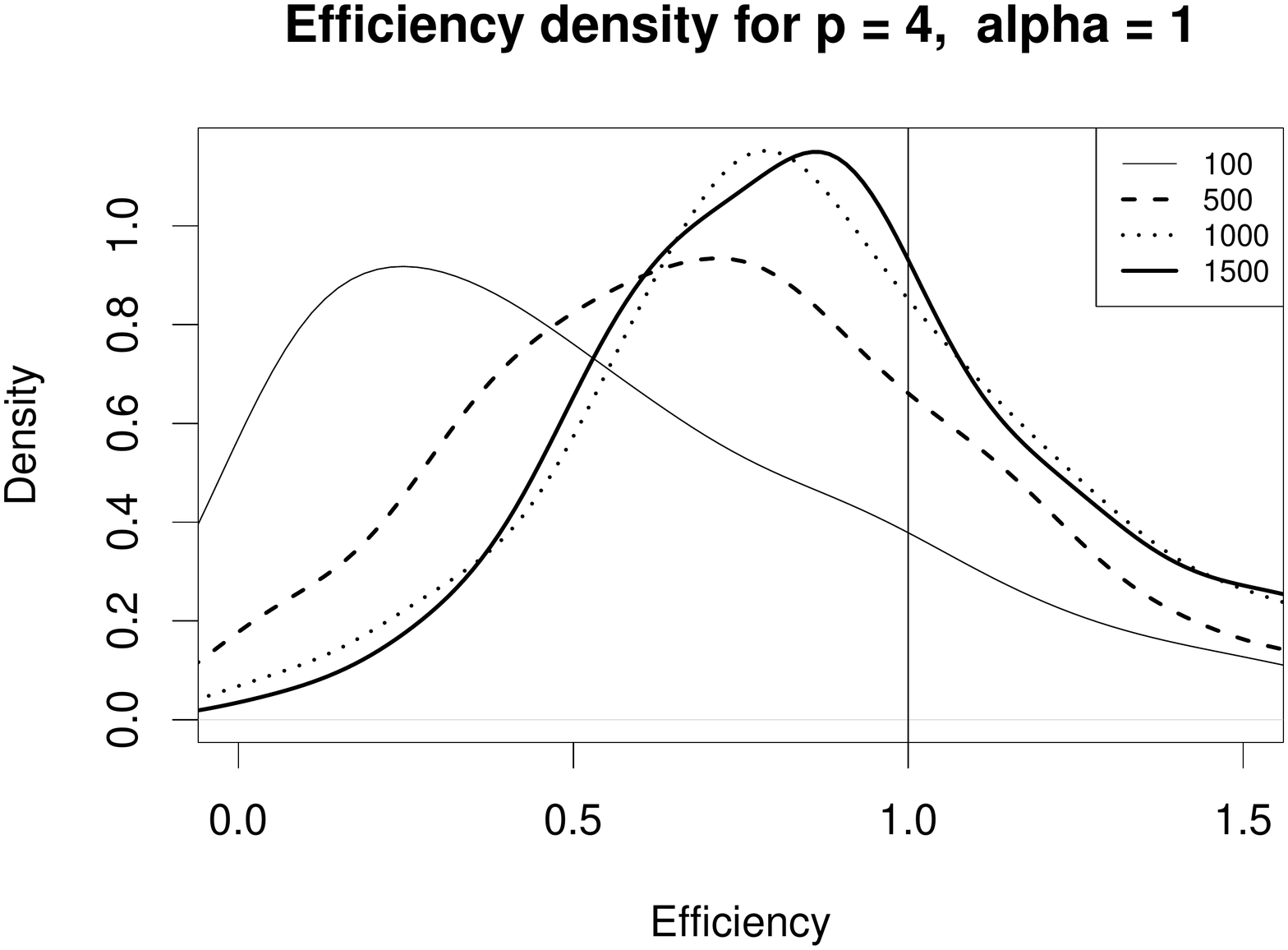}
\end{subfigure}
\begin{subfigure}[b]{0.49\textwidth}
\centering
\includegraphics[width=\textwidth]{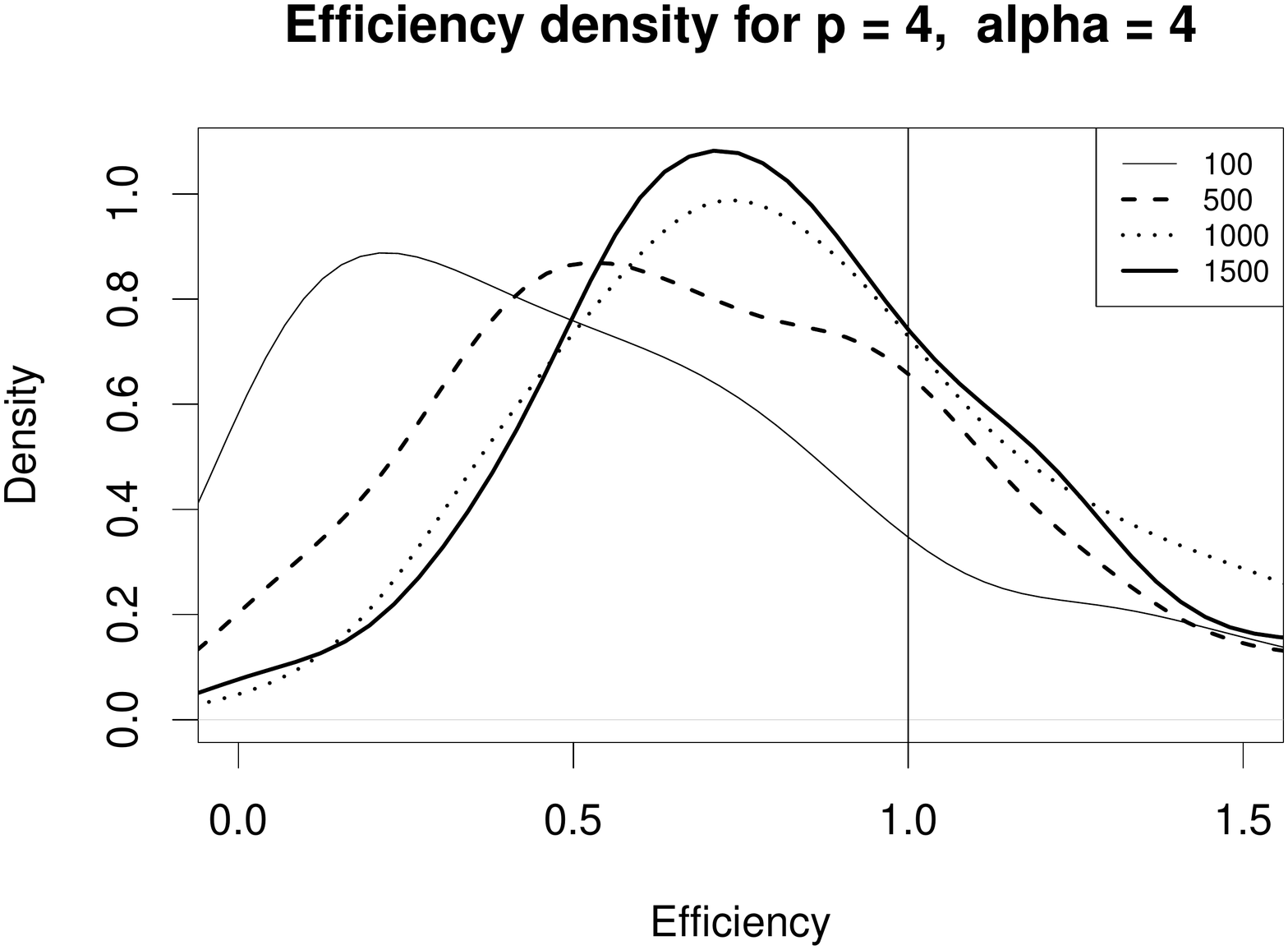}
\end{subfigure}
\begin{subfigure}[b]{0.49\textwidth}
\centering
\includegraphics[width=\textwidth]{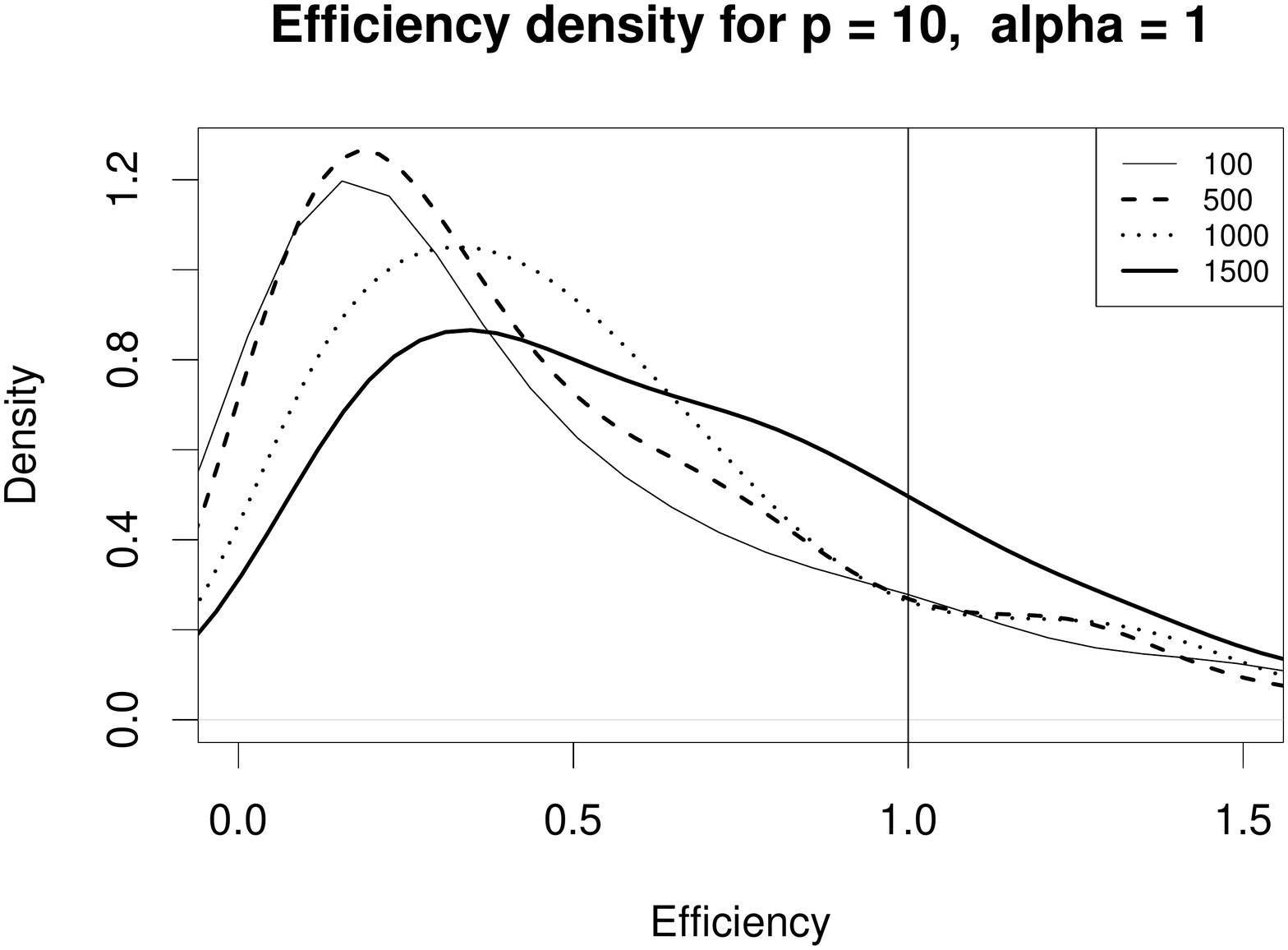}
\end{subfigure}
\begin{subfigure}[b]{0.49\textwidth}
\centering
\includegraphics[width=\textwidth]{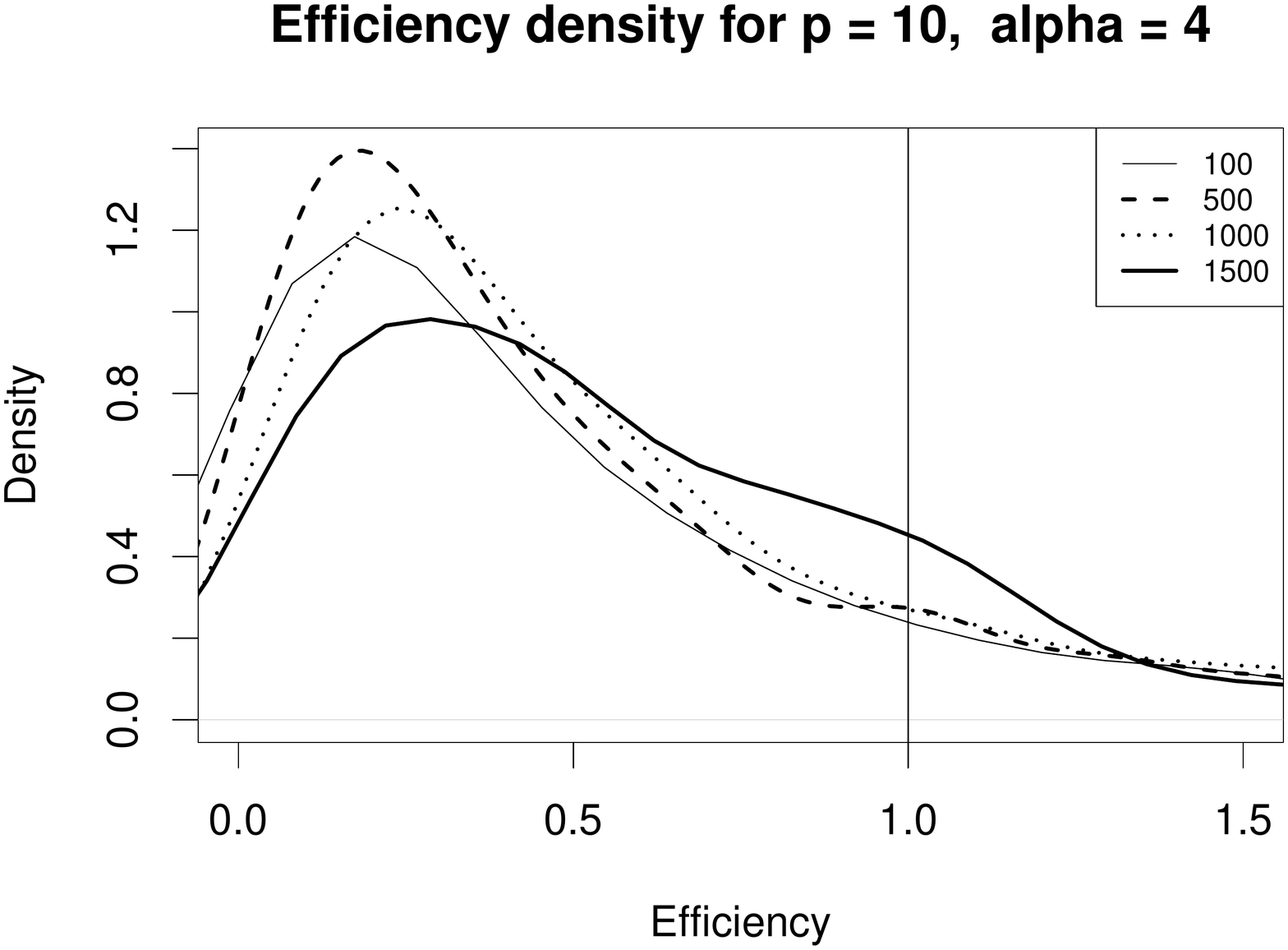}
\end{subfigure}
\caption{Estimated distributions of relative efficiency between $\tilde{m}_{SBK,\alpha}\left(u\right)$ and $\tilde{m}_{O,\alpha}\left(u\right)$.}
\label{figure:eff}
\end{figure}
In Figure \ref{figure:eff}, the densities of relative efficiency distributions for n = 100, 500, 1000, 1500, d = 4, 10 are presented. It is clear to see the relative efficiencies are converging to 1 for low and high dimensions. However, for high dimensions, the converge rate is slower than in low dimensions.

\section{Application}
\label{sec:app}
\begin{figure}[H]
\centering
\begin{subfigure}[b]{0.49\textwidth}
\centering
\includegraphics[width=\textwidth]{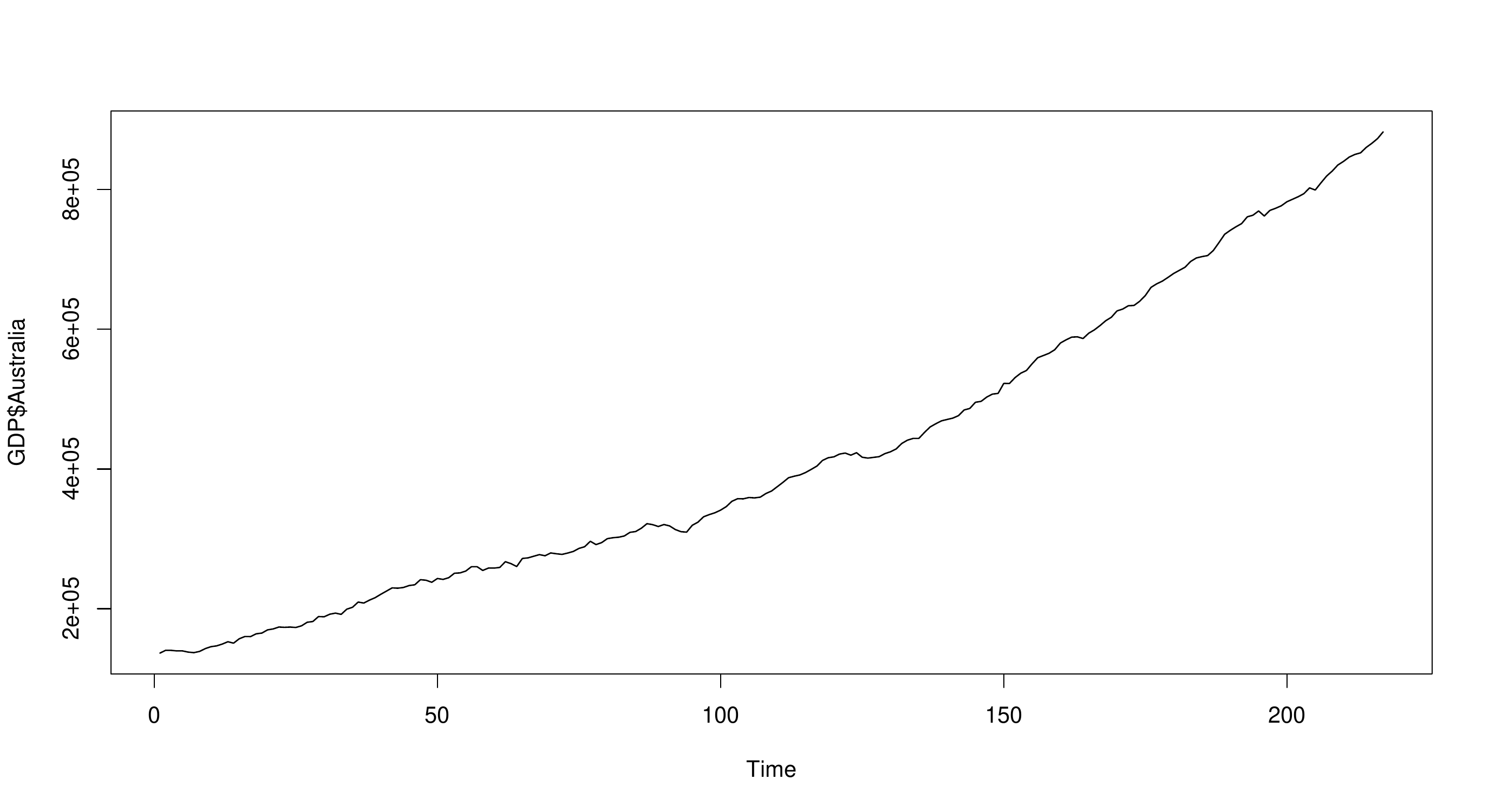}
\subcaption{Quarterly GDP}
\label{figure:gdp}
\end{subfigure}
\begin{subfigure}[b]{0.49\textwidth}
\centering
\includegraphics[width=\textwidth]{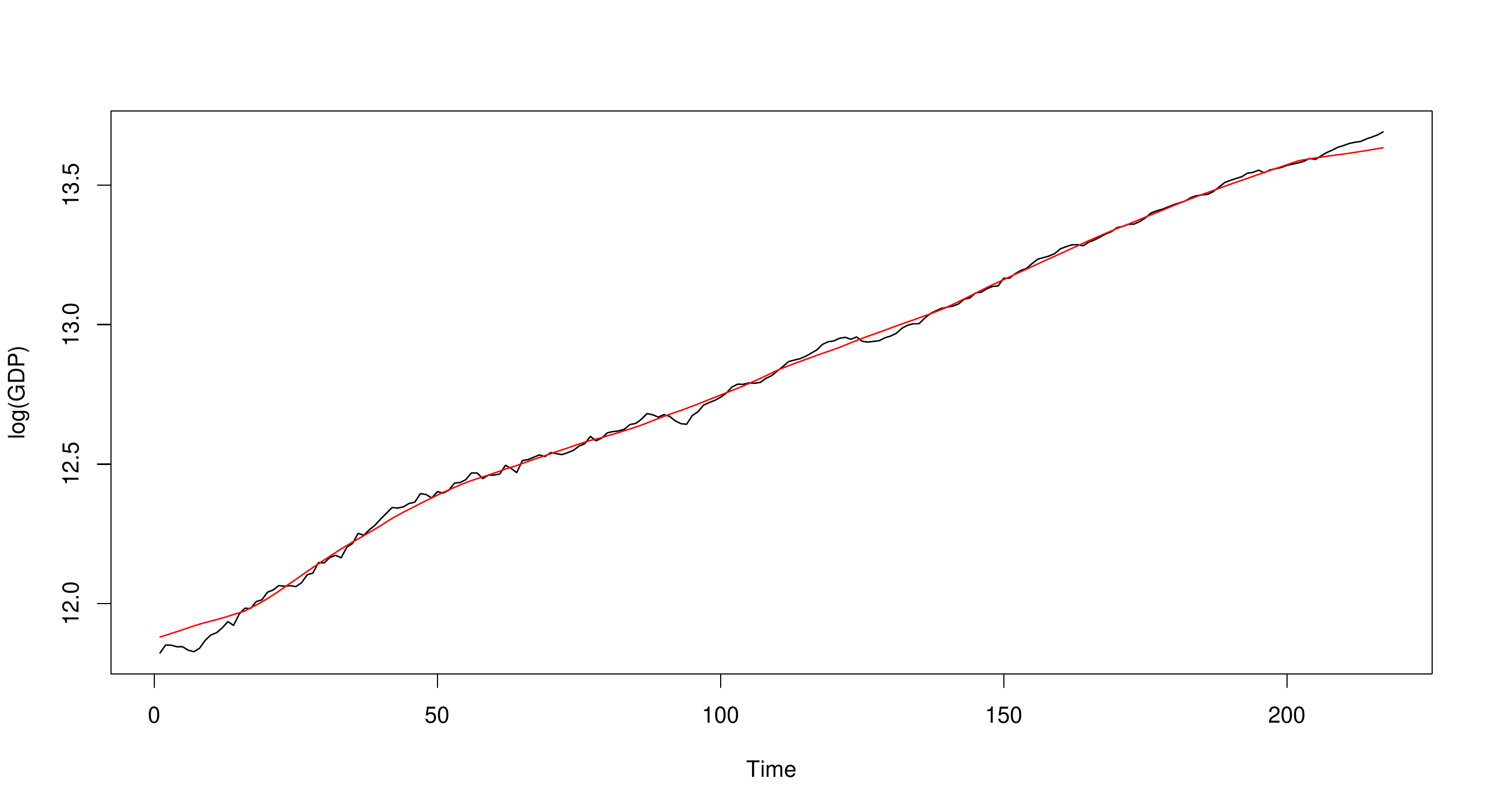}
\subcaption{Log of Quarterly GDP}
\label{figure:loggdp}
\end{subfigure}
\begin{subfigure}[b]{0.49\textwidth}
\centering
\includegraphics[width=\textwidth]{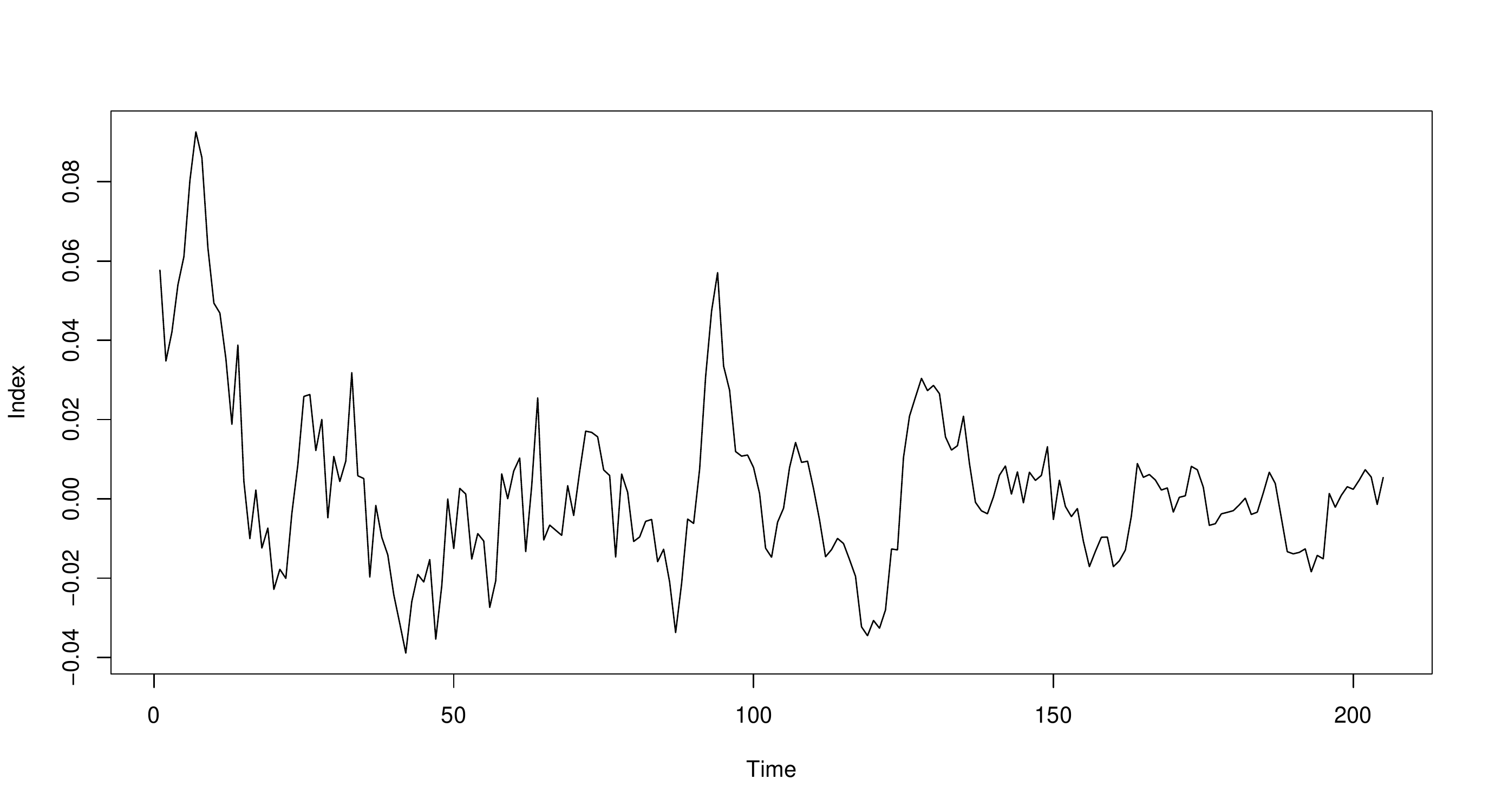}
\subcaption{Detrended}
\label{figure:detrend1}
\end{subfigure}
\begin{subfigure}[b]{0.49\textwidth}
\centering
\includegraphics[width=\textwidth]{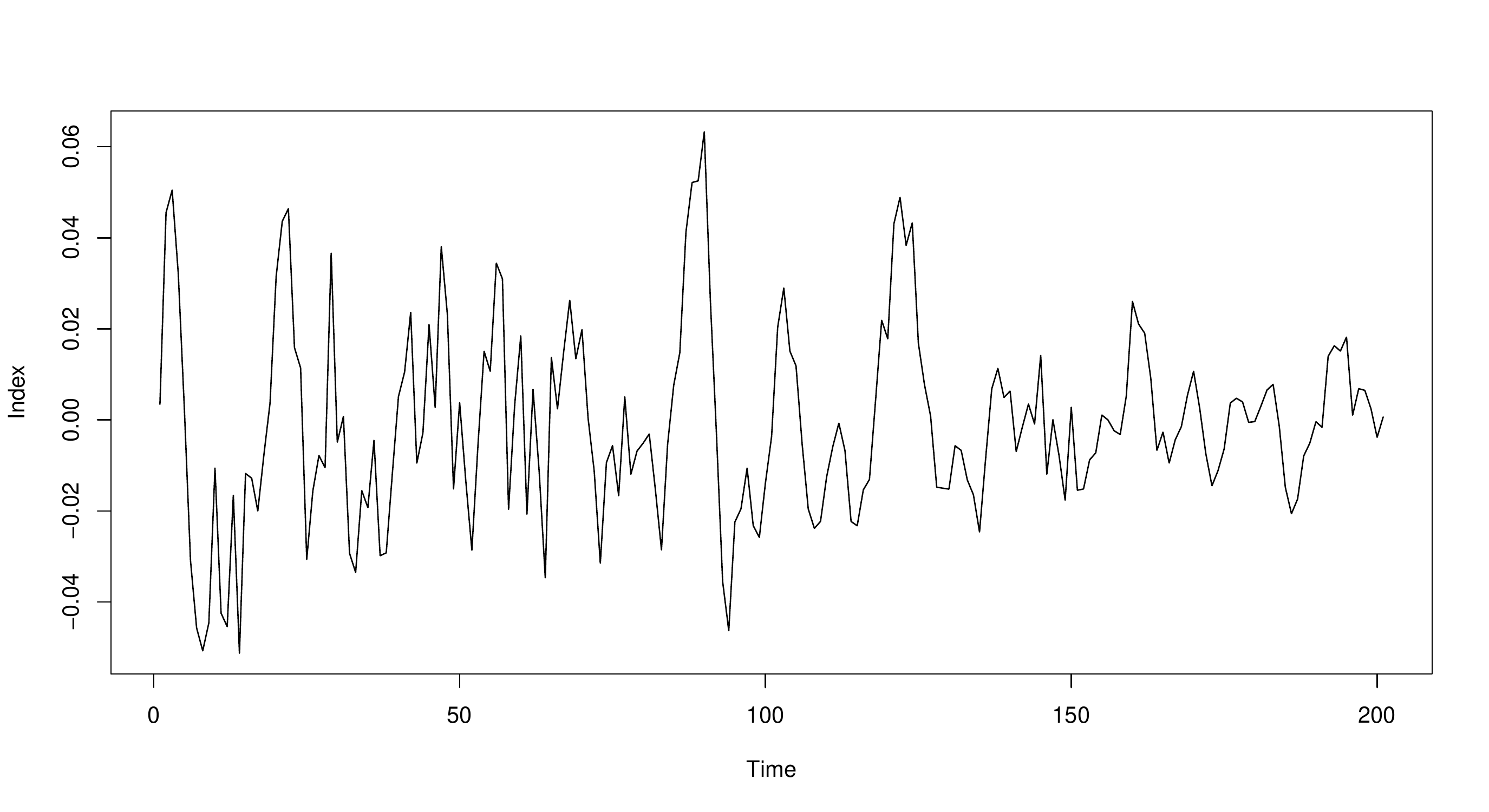}
\subcaption{Differenced-detrended}
\label{figure:detrend2}
\end{subfigure}
\caption{Time series plots for the data: (a) and (b) raw, (c) detrended, (d) differenced and detrended.}
\label{figure:GDP vs Time}
\end{figure}

In this section, we apply our method to the Australia Quarterly GDP data which is obtained from \textit{http://stats.oecd.org}. The data set contains 217 quarterly Australia GDP indices from  Q1 of 1960 to Q1 of 2014 as shown in Figure \ref{figure:gdp}. Usually for economics studies such as GDP, it is better to take the natural log of the data because it will show differnces more clearly. In order to make the series stationary in the mean, we first need to detrend the data. By choosing the bandwith for kernel smoothing to be 30, we fit a line to the data which can be seen in Figure \ref{figure:loggdp}. To make the series stationary in variance, we took the fourth difference due to the data being quarterly. The detrended time series and the differenced-detrended time series are shown in Figures \ref{figure:detrend1} and \ref{figure:detrend2}. 

Recall for our FCAR model in \eqref{eq:fcar}, we need to obtain the delay index $d$ of our delay variable $U_t = X_{t-d}$ and dimension $p$ of $\sum_{\alpha=1}^{p}m_{\alpha}\left(U_{t}\right)X_{t-\alpha}$. Therefore, we estimated the time series with different combination of $d$ and $p$ where $d=\{1,2,...,10\}$ and $p=\{2,3,...,10\}$. Then, the combinition of $d$ and $p$ with minimum MSE will be choosen to be our estimation parameters. The results are shown in Table \ref{table:msetable}. The minimum MSE occurs when $d=7$ and $p=2$.

\begin{table}[H]
\footnotesize
\centering
\begin{tabular}{r| ccccccccc }
\hline
$d/p$ & 2 & 3 & 4 & 5 & 6 & 7 & 8 & 9 & 10        \\
\hline
1&0.000130&0.000638&0.000155&0.000165&0.000168&0.000163&0.000165&0.000168&0.000153\\
2&0.000213&0.000152&0.000239&0.000188&0.000187&0.000182&0.000185&0.000187&0.000178\\
3&0.000132&0.000151&0.000177&0.000184&0.000198&0.000171&0.000170&0.000181&0.000171\\
4&0.000127&0.000150&0.000174&0.000194&0.000176&0.000185&0.000160&0.000157&0.000395\\
5&0.000130&0.000150&0.000179&0.000185&0.000163&0.000149&0.000134&0.000187&0.000171\\
6&0.000144&0.000144&0.000185&0.000184&0.000168&0.000202&0.000176&0.000165&0.000185\\
7&\textbf{0.000116}&0.000126&0.000153&0.000160&0.000174&0.000183&0.000172&0.000173&0.000159\\
8&0.000160&0.000135&0.000152&0.000165&0.000161&0.000171&0.000163&0.000154&0.000156\\
9&0.000118&0.000138&0.000182&0.000163&0.000175&0.000178&0.000178&0.000162&0.000163\\
10&0.000123&0.000130&0.000151&0.000155&0.000166&0.000171&0.000159&0.000167&0.000163\\
\hline
\end{tabular}
\caption{MSE of SBK estimations with different combinitions of $d$ and $p$.}
\label{table:msetable}
\end{table}

\begin{figure}[H]
\centering
\begin{subfigure}[b]{0.49\textwidth}
\centering
\includegraphics[width=\textwidth]{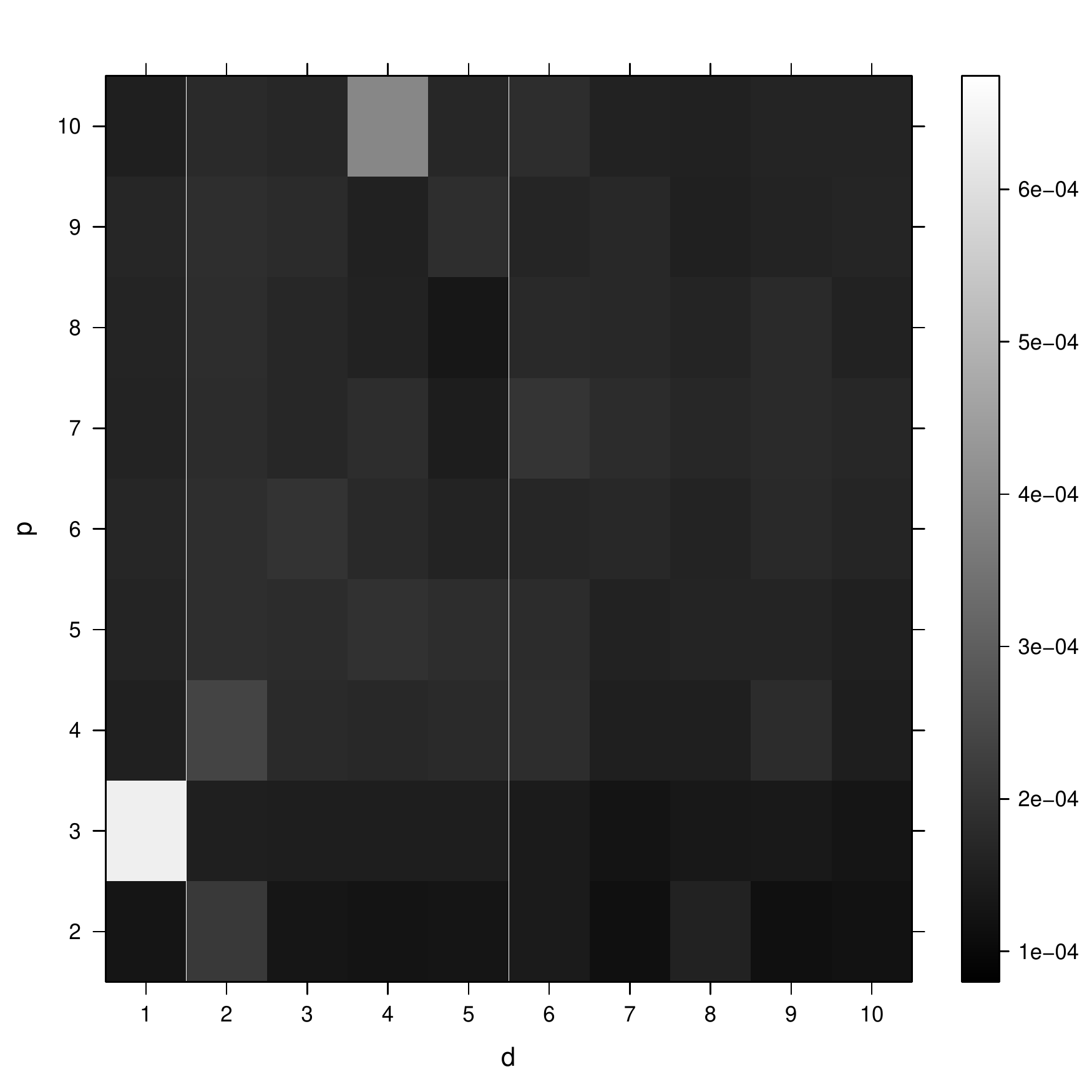}
\subcaption{levels of MSE}
\label{figure:mseall}
\end{subfigure}
\begin{subfigure}[b]{0.49\textwidth}
\centering
\includegraphics[width=\textwidth]{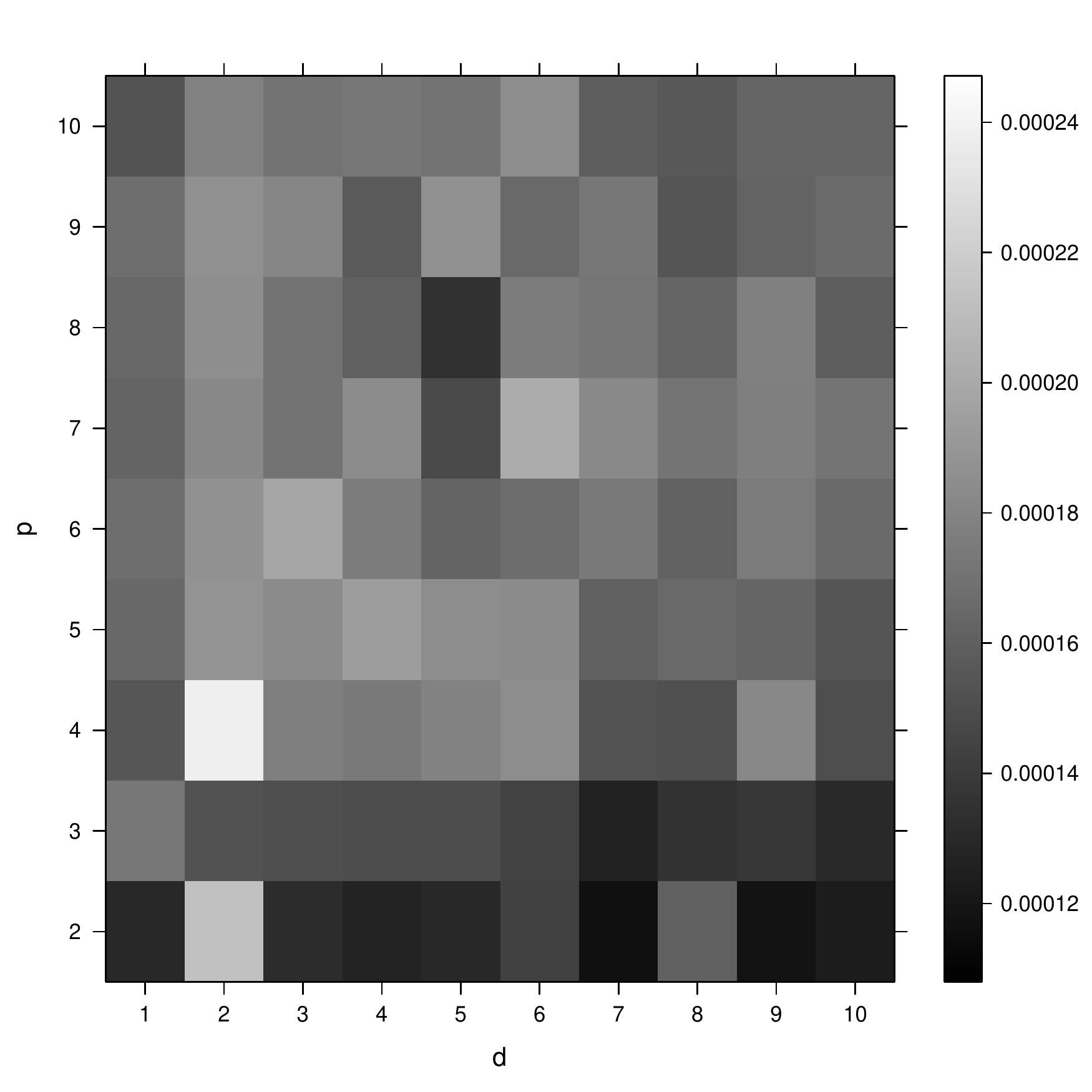}
\subcaption{levels of MSE without extreme values}
\label{figure:msenoextreme}
\end{subfigure}
\caption{MSE of SBK estimations with different combinitions of $d$ and $p$.}
\label{figure:mseplot}
\end{figure}

If we look at the levelplot in Figure \ref{figure:mseplot}, it is easier to see how the MSE values are distributed. The darker the color is, the lower the MSE of SBK estimation is. In \ref{figure:mseall}, even though there are two extremely high MSE values, we can still see that low MSE values generally occur when $p$ is small. Moreover, when we replace the extreme values with the average, we can see a better trend. In \ref{figure:msenoextreme}, we can see that for this perticuclar FCAR model, low MSE values tend to occur when $p$ is small and $d$ is large. 

Since $d=7$ and $p=2$ gives the minimum MSE value when estimating the FCAR model with SBK method, then the FCAR model for this data is 
\[
X_{t}=m_{1}\left(X_{t-7}\right)X_{t-1}+m_{2}\left(X_{t-7}\right)X_{t-2}+\sigma\left(\textbf{X}_{t}\right)\varepsilon_{t}.
\]
To compare our SBK method, we estimated the data again with the ARIMA model with order 1,
\[
X_{t}= c + \psi X_{t-1}+\varepsilon_{t}
\]
where $c=0.0042$ and $\psi=0.8776$.

\begin{figure}[H]
\centering
\includegraphics[width=150mm]{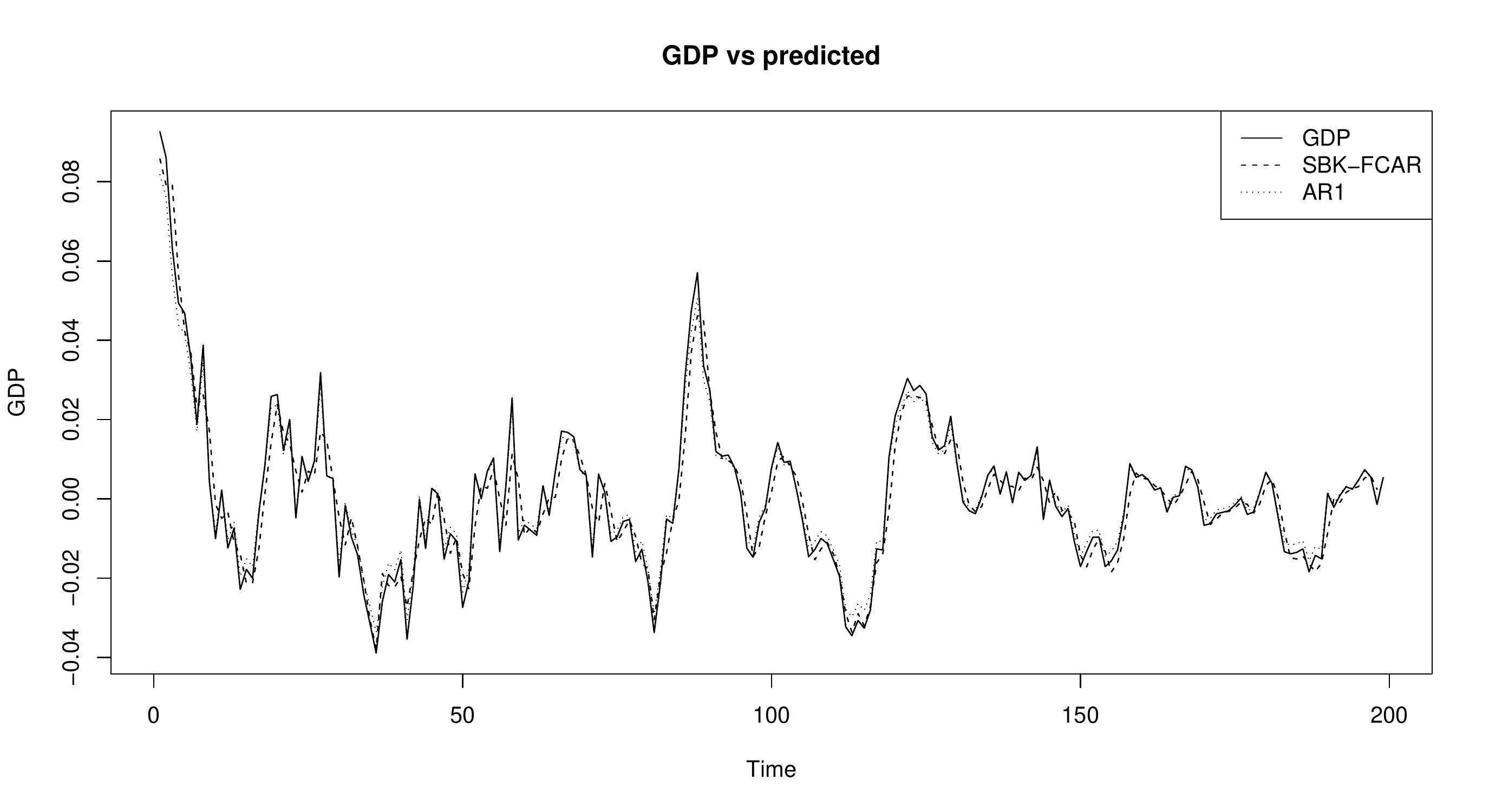}
\caption{SBK on FCAR estimation and ARIMA estimation.}
\label{figure:estimated}
\end{figure}
\begin{figure}[H]
\centering
\begin{subfigure}[b]{0.49\textwidth}
\centering
\includegraphics[width=\textwidth]{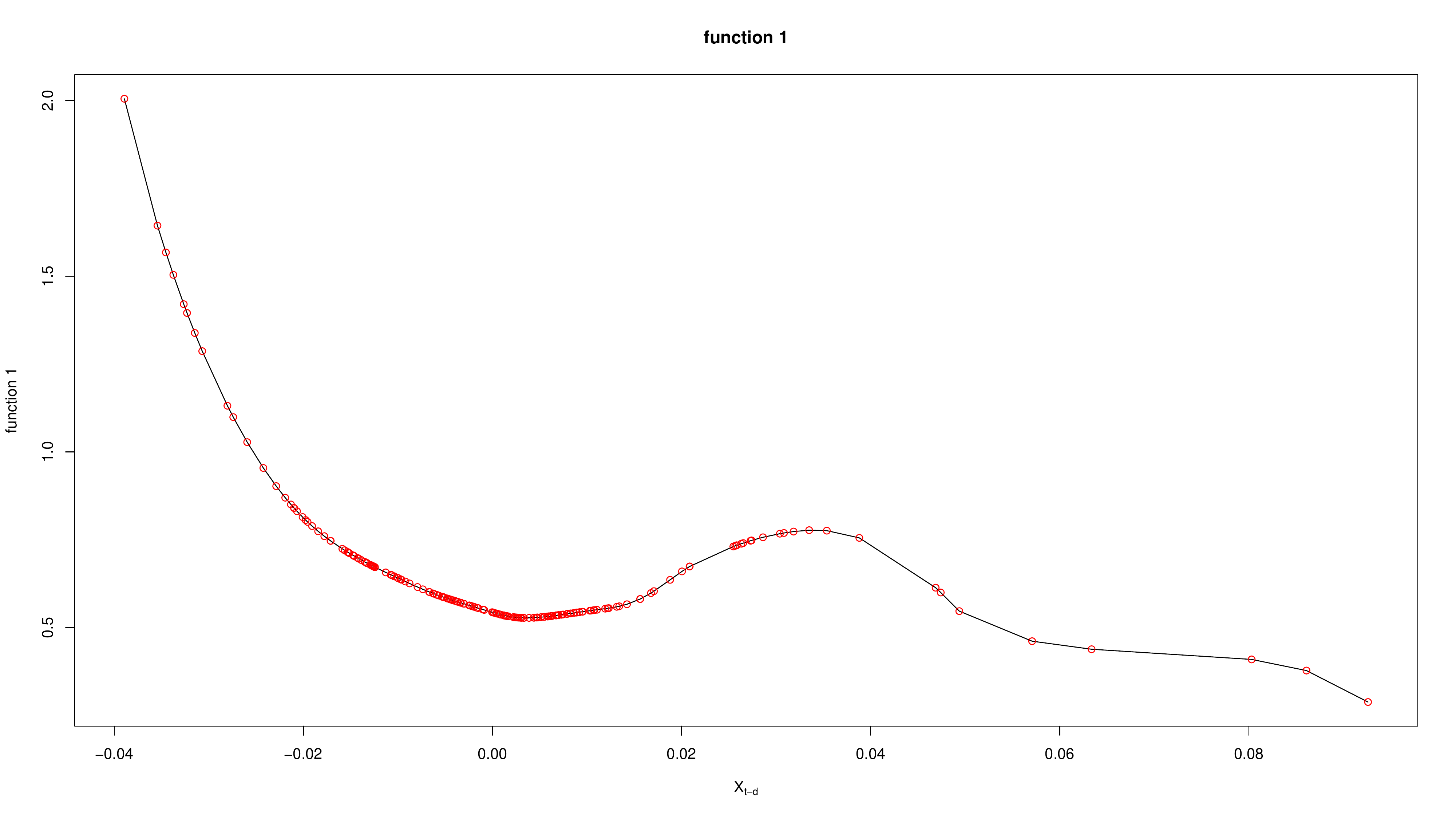}
\end{subfigure}
\begin{subfigure}[b]{0.49\textwidth}
\centering
\includegraphics[width=\textwidth]{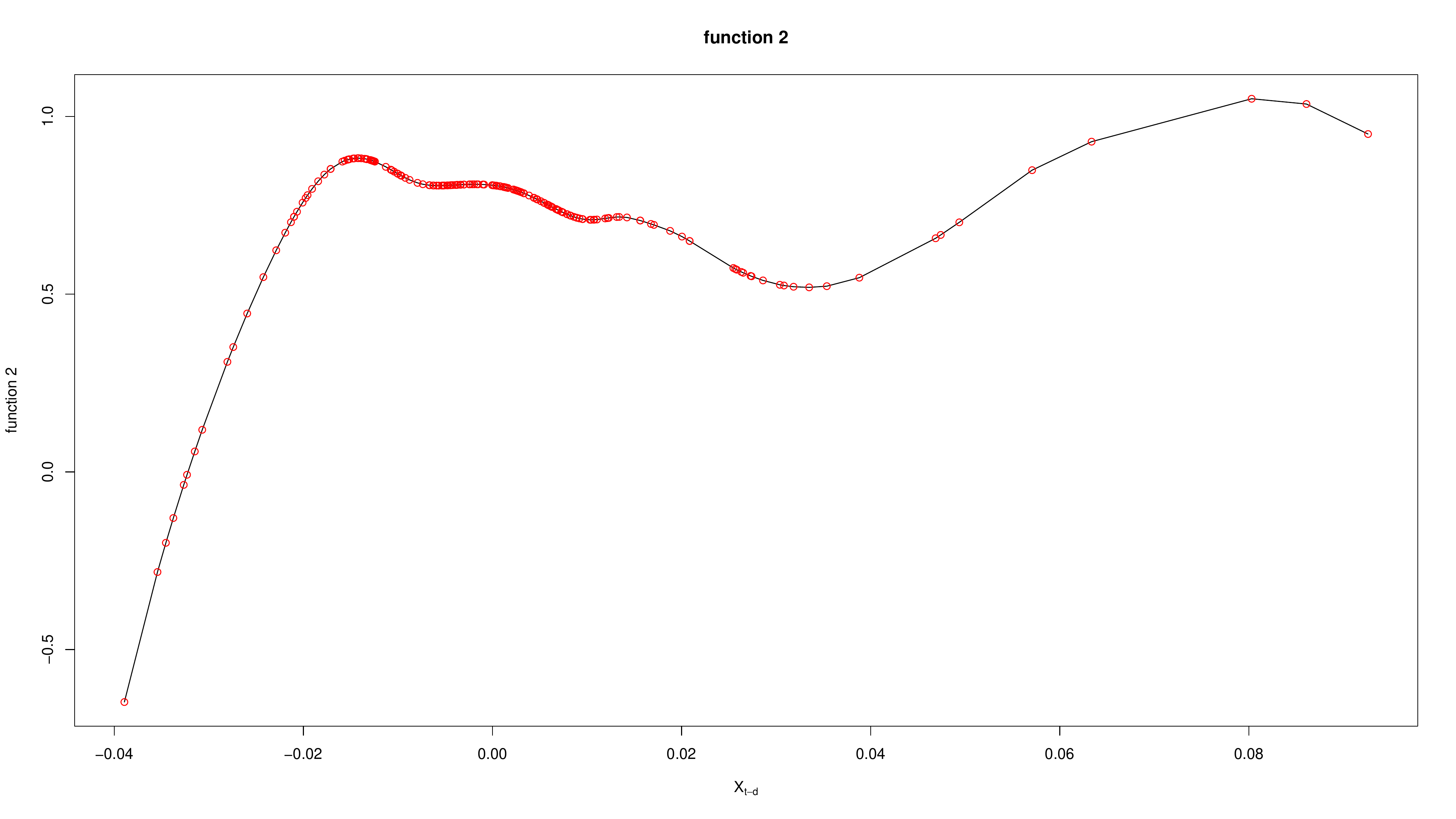}
\end{subfigure}
\caption{Functional Coefficients of our FCAR Model}
\label{figure:m}
\end{figure}

The MSE values for both methods are small. For SBK on FCAR model, MSE is 0.0001164. For ARIMA model, MSE is 0.0001104. Therefore, our SBK estimation on the FCAR model is a satisfactory fit. In Figure \ref{figure:estimated}, we see that both methods fitted the function well. The two coefficient functions we obtained of our FCAR model are shown in Figure \ref{figure:m}. These functions may be used to help professional economists interpret the data. 

\section{Conclusion}
\label{sec:con}
In this paper we have discussed the Spline Backfitted Kernel (SBK) method to estimate the Functional-Coefficient Autoregressive (FCAR) models. This method is breaking a $p$-dimenional problem into $p$-univariate problems, reducing the ``curse of dimensionality." 
This is achieved by first ``pre-estimating" all component functions other than the function of interest with splines, then the difference between observed time series and sum of pre-estimates are used as pseudo-responses for kernel smoothing to estimate the function of interest. 

We showed this method is oracally efficient like local linear estimations in one dimension. Moreover, the speed of this procedure is very faster. One hundred replications with order 10 for sample sizes $n=100, 500, 1000, 1500$ took about 40 minutes on a Macbook Air with a 1.4 GHz Intel Core i5 processor and 4GB RAM. The combination of fast computational speed and asymptotic accuracy for high dimension regression is very appealing.

\newpage
\bibliographystyle{model2-names}
\bibliography{references}

\end{document}